\documentclass[12pt]{article}

\pdfoutput=1

\usepackage{amsmath,amssymb,amsfonts,amscd,mathrsfs}
\usepackage{xcolor}
\definecolor{darkblue}{rgb}{0.1,0.1,.7}
\usepackage[colorlinks, linkcolor=darkblue, citecolor=darkblue, urlcolor=darkblue, linktocpage]{hyperref} 
\usepackage[square, comma, compress,numbers]{natbib}
\usepackage[]{graphicx}
\usepackage{geometry}
\usepackage[margin=10pt,font=small,labelfont=bf]{caption}

\newcommand{\expec}[1]{\langle #1 \rangle}

\newcommand{\mca}[1]{{\mathcal #1}}

\newcommand{\pd}[0]{\partial}

\newcommand{\DD}[0]{\Delta}

\def\l{l} 

\def\n{\nu}

\newcommand{\reef}[1]{(\ref{#1})}

\newcommand{\eps}{\epsilon}
\def\beq{\begin{equation}} 
\def\eeq{\end{equation}} 
\def\del {\partial} 
\def\nn{\nonumber} 
 
\def\bR {\mathbb{R}} 
\def\bC {\mathbb{C}} 
\def\calO {{\cal O}} 
 
\def\calD {{\cal D}} 
\def\calF {{\cal F}} 
 
\def\calP {{\cal P}} 
\def\bn{{\mathbf{n}}}
 
\def\half{\textstyle\frac 12}
\def\ge{\geqslant}
\def\le{\leqslant}
\def\geq{\geqslant}
\def\leq{\leqslant}

\numberwithin{equation}{section}
\begin{document}

\vspace*{-.6in} \thispagestyle{empty}
\begin{flushright}
LPTENS--13/05\\
CERN-PH-TH/2013-043
\end{flushright}
\vspace{1cm} {\Large
\begin{center}
{\bf Radial Coordinates for Conformal Blocks}\\
\end{center}}
\vspace{1cm}
\begin{center}
{\bf Matthijs Hogervorst$^{a,b}$, Slava Rychkov$^{b,a,c}$}\\[2cm] 
{
${}^a$ Laboratoire de Physique Th\'{e}orique de l'\'{E}cole normale sup\'{e}rieure, Paris, France\\
${}^b$ CERN, Theory Division, Geneva, Switzerland\\
${}^c$ Facult\'{e} de Physique, Universit\'{e} Pierre et Marie Curie, Paris, France}

\end{center}

\vspace{4mm}

\begin{abstract}
We develop the theory of conformal blocks in CFT${}_d$ expressing them as power series with Gegenbauer polynomial coefficients. Such series have a clear physical meaning when the conformal block is analyzed in radial quantization: individual terms describe contributions of descendants of a given spin. Convergence of these series can be optimized by a judicious choice of the radial quantization origin. We argue that the best choice is to insert the operators symmetrically. We analyze in detail the resulting ``$\rho$-series" and show that it converges much more rapidly than for the commonly used variable $z$.
We discuss how these conformal block representations can be used in the conformal bootstrap. In particular, we use them to derive analytically some bootstrap bounds whose existence was previously found numerically.
\end{abstract}
\vspace{.2in}
\vspace{.3in}
\hspace{0.7cm} March 2013

\newpage

\tableofcontents 

\section{Introduction}
\begin{figure}[htbp]
\begin{center}
\includegraphics[scale=0.4]{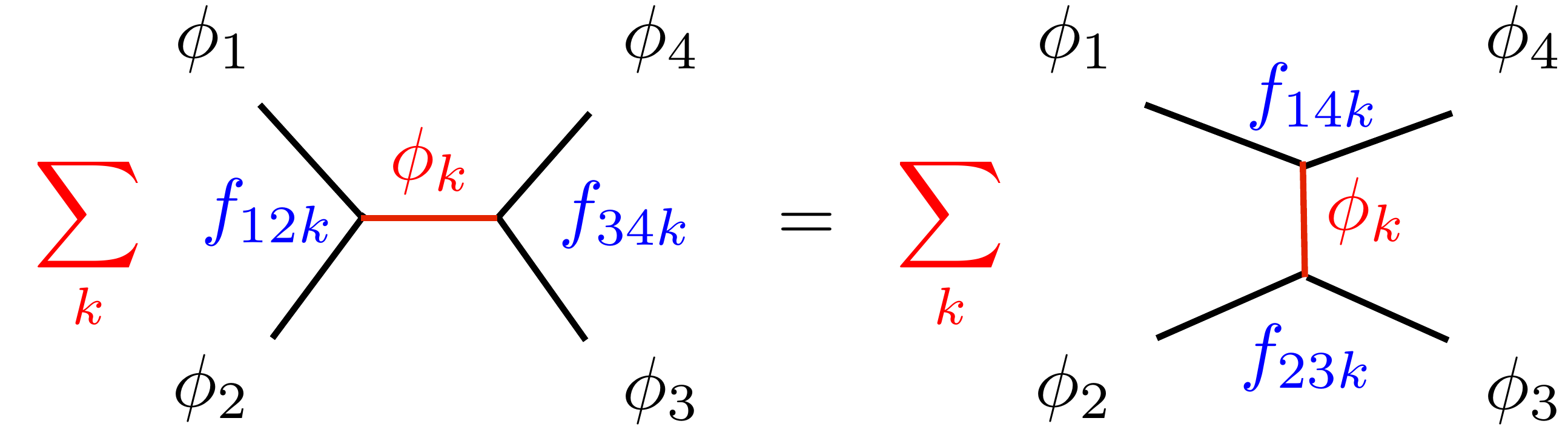}
\caption{Any CFT is characterized by conformal data---primary operator dimensions and spins $(\Delta_i,\l_i)$ and the OPE coefficients $f_{ijk}$. Using the OPE, the four point functions can be expanded into conformal partial waves, fixed by conformal symmetry in terms of the operator quantum numbers, times the products of the OPE coefficients. That the different expansions agree is a nontrivial constraint on the conformal data.}
\label{fig:ass}
\end{center}
\end{figure}

Recent years have seen a revival of the bootstrap approach to Conformal Field Theory (CFT) in higher dimensions \cite{Rattazzi:2008pe,Rychkov:2009ij,Heemskerk:2009pn,Caracciolo:2009bx,Poland:2010wg,Rattazzi:2010gj,
Rattazzi:2010yc,Vichi:2011ux,Poland:2011ey,ElShowk:2012ht,Liendo:2012hy,ElShowk:2012hu,
Fitzpatrick:2012yx,Komargodski:2012ek}. Recall that conformal bootstrap aims to control a CFT by imposing the associativity constraint on the operator algebra. In practice this is done by taking a correlation function of four primary operators\footnote{These are called quasi-primaries in $d=2$ dimensional CFT.}, expanding it into conformal partial waves, and demanding that the different channels agree (see Fig.~\ref{fig:ass}). A central role in this program is played by conformal blocks---functions of the cross ratios obtained by stripping the conformal partial waves from the trivial $x$-dependent factors. 

The theory of conformal blocks in dimension $d\ge 3$ was started in the 70's \cite{Ferrara:1973vz,Ferrara:1974nf,Ferrara:1974ny,Polyakov:1974gs}, with many recent valuable contributions \cite{DO1,DO2,DO3,Costa:2011dw,ElShowk:2012ht,SimmonsDuffin:2012uy,Osborn:2012vt}.\footnote{In $d=2$ dimensional CFT, one distinguishes the ``small" conformal blocks defined by summing over the $SL(2,\bC)$ descendants and the ``big" blocks defined by summing over the Virasoro descendants. Although our focus is on higher dimensions, the blocks considered here reduce to the ``small" blocks in $d=2$. They can also be viewed as the $c\to\infty$ limits of the ``big" blocks \cite{Zamolodchikov:1985ie}.} Especially the explicit expressions for even $d$ found by Dolan and Osborn \cite{DO1,DO2} were instrumental for the first practical applications of the bootstrap. In general $d$, an approach to evaluating the conformal blocks and their derivatives was developed last year in \cite{ElShowk:2012ht}, and applied in the bootstrap analysis of the 3$d$ Ising model.

In spite of these advances, conformal blocks remain rather mysterious special functions. The purpose of this paper is to demystify them via a concrete and economical approach. We will base our considerations on the fact that the conformal blocks are, first and foremost, sums of contributions of radial quantization states to a matrix element computing a four point function. This point of view is standard in the 2$d$ CFT literature \cite{Belavin:1984vu}. Recently \cite{Pappadopulo:2012jk}, it proved useful in general $d$ to study the convergence rate of the conformal block decomposition. Here we develop it to its logical end.

The paper is organized as follows. In section \ref{sec:z} we introduce the radial quantization representation of conformal blocks.
Here we consider the four point function in the frame where two points are fixed at $0$ and $\infty$. The expansion parameter in this frame coincides with the Dolan-Osborn variable $z$. The heart of the paper is section \ref{sec:rho}, where we switch to a different coordinate $\rho$, which corresponds to the frame with points inserted symmetrically with respect to the origin. We demonstrate the advantages of this frame for evaluating the conformal blocks: the expansion parameter $\rho$ is smaller than $z$; the $\rho$-series converges everywhere where the block is expected to be regular; its coefficients are bounded independently of $\Delta$ and $l$. The last two properties are not true for the $z$-series expansions. 

We foresee that the $\rho$-series representations of conformal blocks will find many applications in the conformal bootstrap program; we outline some in section \ref{sec:appl}. An especially neat application is the ``toy bootstrap equation" (section \ref{sec:toy}), by means of which it is possible at last to get analytic understanding of why the methods of \cite{Rattazzi:2008pe} were successful in producing upper bounds on the operator spectrum. 

We conclude in section \ref{sec:concl}. Appendix \ref{sec:largeL} contains the proof of boundedness of the $\rho$-series expansion coefficients.

\section{Conformal blocks in the Dolan-Osborn coordinates}
\label{sec:z}
\subsection{General structure}
\label{sec:zgen}
For simplicity we will focus on the Euclidean-space correlator of four identical scalar primaries.\footnote{The generalization to non-identical scalars is straightforward, and we will comment on the non-scalar case in section \ref{sec:concl}.}  By conformal invariance it has the form
\beq
\expec{\phi(x_1) \phi(x_2) \phi(x_3) \phi(x_4) } = \frac{g(u,v)}{(x^2_{12})^{\Delta_\phi}(x^2_{34})^{\Delta_\phi}}, \qquad x_{ij} \equiv x_i - x_j\,,
\label{eq:fourpt}
\eeq
where $g(u,v)$ is a function of the conformally invariant cross ratios
\beq
u = (x_{12}^2 x_{34}^2)/(x_{13}^2 x_{24}^2), \quad v = ({x_{14}^2 x_{23}^2})/({x_{13}^2 x_{24}^2}).
\label{eq:crossr}
\eeq
The partial wave decomposition of this correlator takes the form:
\beq
\label{eq:g_sum}
g(u,v) =\sum_{\mca{O}} f_{\mca{O}}^2 G_{\mca{O}}(u,v),
\eeq
where the $G_{\mca{O}}(u,v)$ are the conformal blocks of the primary operators appearing in the $\phi\times \phi$ OPE and $f_{\mca{O}}$ are their OPE coefficients. The function $g(u,v)$ computed from this expansion must satisfy the crossing symmetry equation
\beq
v^{\Delta_\phi} g(u,v)=u^{\Delta_\phi} g(v,u)\,,
\label{eq:cross}
\eeq
which imposes constraints on the dimensions, spins, and OPE coefficients $f_\calO$ of the exchanged operators. 
However, our main interest here is not in how to extract these constraints (this will be briefly discussed in section \ref{sec:appl}), but in the conformal blocks themselves.

\begin{figure}[htbp]
\begin{center}
\includegraphics[width=7cm]{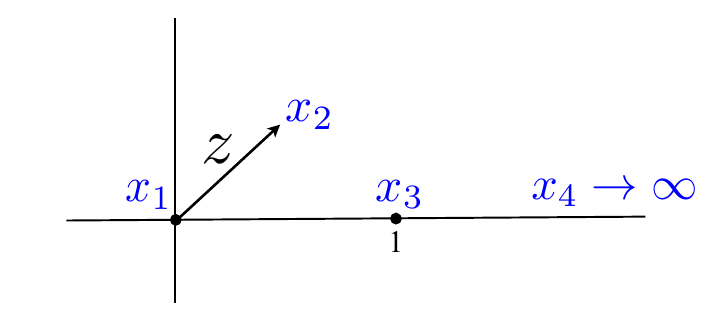}
\caption{By conformal symmetry, three operators can be put at $x_1=0$, $x_3=(1,0,\ldots,0)$, $x_4\to \infty$, with the fourth point  $x_2$ somewhere in the (12)-plane parametrized by the complex coordinate $z$. }
\label{fig:z}
\end{center}
\end{figure}

Starting from the work of Dolan and Osborn \cite{DO1,DO2}, it has become customary to express conformal blocks by changing coordinates from $u,v$ to $z$ and $\bar z\equiv z^*$:
\beq
u=z\bar{z},\qquad v=(1-z)(1-\bar z)\,.
\eeq
The geometrical meaning of the new variables is made clear by assigning three points to $0,1,\infty$ as in Fig.~\ref{fig:z}. 
The complex $z$ is the usual coordinate used in $d=2$ dimensional CFT, but its utility for general $d$ is not a priori obvious. Refs.~\cite{DO1,DO2} discovered that conformal blocks in $d=4$ and in all even dimensions take particularly simple expressions in these coordinates.
Here we will work with any $d$, even or odd.\footnote{In fact, as we will see below, conformal blocks depend analytically on $d$. The conformal bootstrap equation with analytically continued blocks can be formally considered for any $d$. It can be taken as a nonperturbative definition of CFT in fractional dimensions \cite{Rychkov:2011et}.}

To avoid possible misunderstanding, we should stress that although we parametrize the conformal blocks by a complex variable $z$, we never use complex analysis. Only in the 2$d$ case do the conformal blocks factorize as a holomorphic times antiholomorphic function. For general $d$ considered here, we will treat conformal blocks as smooth real functions in the $z$ plane; 
see the end of this section and section \ref{sec:Zam} for more details.
 
\begin{figure}[htbp]
\begin{center}
\includegraphics[scale=1]{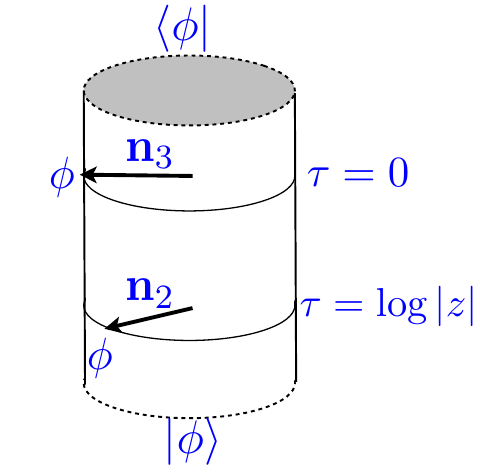}
\caption{Using a Weyl transformation, the configuration in Fig.~\ref{fig:z} is mapped onto a cylinder matrix element with operators inserted as shown.}
\label{fig:radz}
\end{center}
\end{figure}

To exhibit the general structure of conformal blocks in the $z,\bar z$ variables, let's use radial quantization. It is convenient, although not strictly necessary, to do a Weyl transformation which maps the CFT from $\bR^d$ to the cylinder $\bR\times S^{d-1}$. In polar coordinates, the mapping is simply $(r,\bn)\to (\tau,\bn)$, with the cylinder time $\tau=\log r$.
The flat space four point function with points assigned as in Fig.~\ref{fig:radz} then maps to the cylinder matrix element
\beq
\langle \phi|\phi(\tau_3,\bn_3) \phi(\tau_2,\bn_2) |\phi\rangle\,.
\label{eq:cylmat}
\eeq
The operators inserted at zero and infinity map to the radial quantization in- and out-states $|\phi\rangle$ and $\langle \phi|$. The other two insertions are at the cylinder times $\tau_2=\log|z|$ and $\tau_3=0$. We keep both unit vectors $\bn_2$ and $\bn_3$ explicit for future use, but the only rotationally invariant parameter is their scalar product
\beq
\bn_2\cdot \bn_3=\cos\theta,\quad\theta=\arg z\,.
\eeq

The next step is to express \reef{eq:cylmat} by inserting a complete basis of energy eigenstates on $S^{d-1}$. This gives\footnote{The extra factors in the RHS of this formula which follow from the denominator of Eq.~\reef{eq:fourpt}, from the Weyl transformation of operators, and from acting with $\exp(\pm H\tau)$ on the in- and out-states when shifting the operator insertion times to zero, cancel each other; see \cite{Pappadopulo:2012jk} for a more detailed derivation.}
\beq
g(u,v)=\sum _{E} |z|^E  \langle \phi|\phi(0,
\bn_3)|E\rangle\,
\langle E|\phi(0,
\bn_2) |\phi\rangle
\label{eq:cylmat1}
\eeq
 where we took into account that propagating a state of energy\footnote{The presence of the Casimir energy on the sphere, nonzero for even $d$, can be ignored here. This is because we are discussing correlation functions and not the partition function, and so the relevant energy is the one defined subtracting the energy of the ground state.} $E$ for the Euclidean time distance $\tau_3-\tau_2$ will give rise to the factor
 \beq
 e^{-E(\tau_3-\tau_2)}=|z|^E\,.
 \eeq
 
The exchanged states on the sphere are in one-to-one correspondence with the local operators appearing in the OPE $\phi\times\phi$. For the moment we do not distinguish between the primary and descendant states. Every state will come in a multiplet of $SO(d)$. In fact, only symmetric traceless tensor multiplets of spin $j\ge 0$ can couple for the considered correlator.\footnote{One cannot construct an antisymmetric tensor out of a single vector $\bn$, and so the corresponding matrix elements necessarily vanish.} The right matrix element
\beq
\langle E,\{\mu_1,\ldots,\mu_j\} |\phi(0,
\bn_2) |\phi\rangle
\eeq
must be a rank-$j$ symmetric traceless tensor constructed out of the vector $\bn_2$, which is fixed up to a constant:
\beq
\bn_2^{\mu_1}\bn_2^{\mu_2}\ldots \bn_2^{\mu_j}-\text{traces}\,.
\eeq
Analogously, the left matrix element is fixed up to a constant, and so a general term in \reef{eq:cylmat1} will be proportional 
to\footnote{This contraction formula follows from the theory of spherical harmonics and harmonic polynomials, see \cite{Bateman2}, Section 11.2, Lemma 1, and \cite{SteinWeiss}, Chapter 4. 
} 
\beq
(\bn_2^{\mu_1}\bn_2^{\mu_2}\ldots \bn_2^{\mu_j}-\text{traces})(\bn_3^{\mu_1}\bn_3^{\mu_2}\ldots \bn_3^{\mu_j}-\text{traces}) 
\propto C_j^{\n}(\bn_2\cdot \bn_3), \quad \n \equiv d/2-1,
\label{eq:contraction}
\eeq
where $C_j^{\n}$ are the Gegenbauer polynomials. For the integer dimensions of interest $d=2,3,4$, they take the form
\begin{align}
\lim_{\nu\to0}\nu^{-1}C_j^\nu(\cos\theta)&=\frac{2}{j}\cos(j\,\theta)\, \qquad (j \geq 1),\nn\\
C_j^{1/2}(\cos\theta)&=P_j(\cos\theta)\,,\\
C_j^{1}(\cos\theta)&=\frac{\sin[(j+1)\theta]}{\sin\theta}\,.\nn
\end{align}
In particular, for $d=3$ we get the Legendre polynomials. 


We conclude that the function $g(u,v)$ appearing in the four point function \reef{eq:fourpt} must have an expansion of the form:
\beq
g(u,v)=1+\sum p_{E,j} |z|^E 
C_j^{\n}(\cos \theta)\,,\qquad p_{E,j}\ge0.
\label{eq:gexp}
\eeq
where the sum is over all local operators of dimension $E$ and spin $j$ appearing in the OPE $\phi\times\phi$.
Although the coefficients $p_{E,j}$ are left undetermined by this argument, we do know that they must be non-negative. This is because for $\bn_2=\bn_3$ the configuration in Fig.~\ref{fig:radz} becomes reflection-positive. The matrix elements in \reef{eq:cylmat1} are then complex conjugates of each other.

\begin{figure}[htb]
\centering
\includegraphics{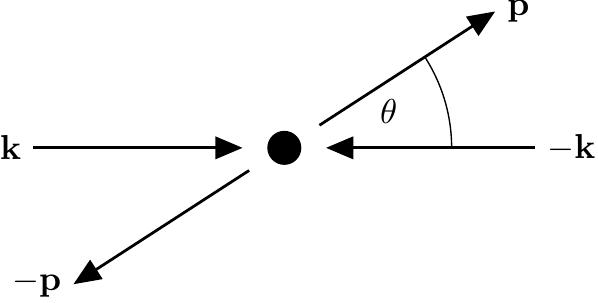}
\caption{\label{fig:scattering}{\small Elastic center-of-mass scattering of two scalar particles. When a spin-$j$ resonance dominates the scattering process, the amplitude is proportional to $P_j(\cos \theta)$.}}
\end{figure}

The appearance of Gegenbauer polynomials in this result is not surprising, as they already arise in the theory of angular momentum in quantum mechanics. When two spinless particles scatter through a spin-$j$ resonance, it is well known that the amplitude is given by the Legendre polynomial of the scattering angle (see Fig.~\ref{fig:scattering}).

Consider now a particular primary operator $\calO$ of dimension $\Delta$ and spin $\l$ occurring in the $\phi\times\phi$ OPE. If we restrict the sum \reef{eq:gexp} to its conformal multiplet, it must represent the conformal block of $\calO$. The conformal multiplet will have descendants of integer-spaced dimensions $\Delta+n$ with spins at level $n$ taking values\footnote{For the short representations some of these spins will not be there.}
\beq
j=\l+n,\l+n-2,\ldots,\max (l-n,l+n\!\!\!\!\mod 2)\,.
\label{eq:j}
\eeq
Moreover, the coefficients $p_{E,j}$ within one conformal multiplet are not independent, since the matrix elements for the descendants will be all proportional to the basic OPE coefficient $f_\calO$. We conclude that the conformal block must have the following expansion:
\beq
G_{\Delta,\l}(u,v)=\sum_{n=0}^\infty |z|^{\Delta+n} \sum_j A_{n,j} \frac{C_j^{\n}(\cos \theta)}
{C_j^\nu(1)}\,,\qquad A_{n,j}\ge0,
\label{eq:Gexp}
\eeq
where the positive coefficients $A_{n,j}$ are some universal functions of $\Delta,\l$, and $d$ that are fixed by conformal symmetry. We normalize the total conformal block by the condition $A_{0,\l}=1$.\footnote{This normalization relates to the one used by Dolan and Osborn in \cite{DO1,DO2} as  $G^{\text{here}}_{\Delta,\l}=(-2)^l (\nu)_\l/(2\nu)_\l G^{\text{there}}_{\Delta,\l}.$}

The Gegenbauer normalization factors,
\beq
C_j^\nu(1)={(2\n)_j}/{j!}\,,
\eeq
are included in \reef{eq:Gexp} for later convenience and also to ensure a smooth $\nu\to0$ limit for $d=2$. Here, $(x)_n = \Gamma(x+n)/\Gamma(x)$ are the Pochhammer symbols.

The formula \reef{eq:Gexp} is the main result of this section. It should be noted that Ref.~\cite{DO2} already used an expansion of conformal blocks into Gegenbauer polynomials, because they turn out to form a convenient basis for solving the Casimir differential equation recursively (see the next section). Ref.~\cite{Fitzpatrick:2012yx}, Eq.~(78), observed that in any number of dimensions conformal blocks can be expanded in $\cos(j\,\theta)$ with positive coefficients. For $d=2$ our result says the same, although for general $d$ our conclusion is stronger. To obtain their result, one runs the above argument classifying states into multiplets with respect to the $SO(2)$ subgroup of $SO(d)$ acting in the (12)-plane. In particular, the Gegenbauer polynomials for any $\nu\ge0$ have positive expansions in $\cos(j'\,\theta)$, $j'\le j$.

The region of convergence of the expansion \reef{eq:Gexp} will be limited to $|z|<1$, which is the condition for the operators $\phi_2$ and $\phi_3$ in \reef{eq:cylmat} to be time-ordered on the cylinder. However, the actual domain $X$ of regularity of the conformal block as a function of $z$ is larger; it is given by the complex plane minus the $(1,+\infty)$ cut along the real axis:
\beq
X=\mathbb{C}\backslash(1,+\infty)\,.
\label{eq:X}
\eeq 
Everywhere in this region the blocks will be real analytic, except at $z=0$ because of the $|z|^\Delta$ factor.
For every point in $X$ one can find a sphere which separates $x_1$ and $x_2$ from $x_3$ and $x_4$. Choosing the center of this sphere as a radial quantization origin, one can prove the regularity of the conformal block for such $z$. The blocks will be singular on the cut, because the separating sphere jumps when $z$ crosses it. In section \ref{sec:rho}
below we will construct expansions convergent in the full region $X$. 
But first we would like to study the coefficients of the expansion \reef{eq:Gexp}.

\subsection{Expansion coefficients from the Casimir equation}

We would like to compute the coefficients $A_{n,j}$  in \reef{eq:Gexp}. In principle, this can be done following the radial quantization method to its logical end: imposing the constraints of conformal invariance in the OPE and evaluating the norms of the descendants. The example of scalar exchanged primaries and their first two descendant levels was considered in \cite{Pappadopulo:2012jk}. However, it is far more efficient to use a different method first proposed in \cite{DO2}.

The idea is that the conformal block satisfies an eigenvalue equation of the form 
\beq
\mca{D} G_{\DD,\l}(u,v) = C_{\DD,\l} \, G_{\DD,\l}(u,v), \qquad C_{\DD,\l} = \DD(\DD-d) + \l(\l+d-2),
\label{eq:Caseq}
\eeq
where $\mca{D}$ is a second-order partial differential operator. To get it, one acts on the four point function with the combination of conformal group generator (in the $SO(d+1,1)$ notation)
\beq
\half (L_{AB}^{(1)}+L_{AB}^{(2)})(L^{(1)AB}+L^{(2)AB}),
\label{eq:M2}
\eeq
where the generators $L^{(i)}$ acts on the operator inserted at $x_i$. By conformal invariance of the OPE, this combination can be pushed through to act as the quadratic Casimir on the operators appearing in the OPE $\phi(x_1)\times\phi(x_2)$. All terms within a given conformal family will have the same Casimir eigenvalue $C_{\DD,\l}$. This gives a differential equation for the conformal partial wave.

In the $z,\bar z$ coordinates the operator $\calD$ takes the form \cite{DO2}
 \beq 
 \half \mathcal{D}=[z^{2}(1-z)\partial_{z}^{2}- z^{2}\partial_{z}]+[\bar{z}^{2}(1-\bar
{z})\partial_{\bar{z}}^{2}-\bar{z}^{2}\partial
_{\bar{z}}]  +2\nu\displaystyle\frac{z\bar{z}}{z-\bar z}\left[  (1-z)\partial
_{z}-(1-\bar{z})\partial_{\bar{z}}\right]\,. 
\label{eq:calD}
\eeq
For our purposes it will be convenient to express it in the coordinates 
\beq
s=|z|,\quad \xi=\cos\theta=(z+\bar z)/(2|z|)\,.
\eeq
We find:
\begin{align}
\calD&=\calD_0+\calD_1, \nn \\
\mca{D}_0&=s^2 \pd_s^2 + (2 \nu +1) \left[\xi \, \pd_\xi -  s\, \pd_s \right] - \left(1-\xi^2\right) \pd_\xi^2, 
\label{eq:D0D1}\\
\mca{D}_1&=s \left[- \xi  s^2 \pd_s^2 +2\left(1-\xi ^2\right) s\, \pd_s \pd_\xi - \xi  s\, \pd_s - \left(2 \nu +\xi ^2\right) \pd_\xi +\xi\left(1 -\xi^2\right) \pd_\xi^2 \right].\nn
\end{align}
The terms are grouped in such a way that $\calD_0$ preserves homogeneity in $s$ while $\calD_1$ increases it by 1.

We now apply this operator to \reef{eq:Gexp}, which we write as
\beq
G_{\Delta,l}=\sum_{n=0}^\infty \sum_j A_{n,j} \calP_{\Delta+n,j},\qquad  \calP_{E,j}(s,\xi)\equiv s^{E} \frac{j!}{(2\n)_j} C_j^{\n}(\xi)\,.
\label{eq:ansatz}
\eeq
Using the properties of Gegenbauer polynomials, it is easy to see that $\calP_{E,j}$ are eigenfunctions of $\calD_0$. The eigenvalue depends on the dimension and spin in the same way as the Casimir:
\beq
\calD_0 \calP_{E,j}=C_{E,j}\calP_{E,j}\,.
\eeq
The $\calD_1$ also acts simply in this basis: 
\begin{align}
\calD_1 \calP_{E,j}&= -\gamma^+_{E,j} \calP_{E+1,j+1} -\gamma^-_{E,j}\calP_{E+1,j-1}, \,\nn\\
\gamma^+_{E,j}&= \frac{(E+j)^2 (j+2\nu)}{2(j+\nu)}, \qquad \gamma^-_{E,j}= \frac{(E-j-2\nu)^2 j}{2(j+\nu)}.
\label{eq:D1act}
\end{align}
Applying these formulas, equation \reef{eq:Caseq} can be solved order by order in $s$. We find that the coefficients $A_{n,j}$ must satisfy the following recursion relation:
\beq
\left( C_{\DD+n,j} - C_{\DD,\l} \right) A_{n,j}= \gamma^+_{\DD+n-1,j-1} A_{n-1,j-1} + \gamma^{-}_{\DD+n-1,j+1} A_{n-1,j+1}.
\label{eq:recursion}
\eeq
Starting from the initial conditions 
\beq
A_{0,j} = \delta_{j \l}
\label{eq:init}
\eeq
this recursion determines all coefficients $A_{n,j}$. One can check that 
\beq
C_{\DD+n,j} - C_{\DD,\l}>0
\eeq
if $\Delta$ satisfies unitarity bounds and $j$ is in the range \reef{eq:j}. So the coefficients generated by the recursion are manifestly positive, in agreement with the previous section.

For illustration, here is what the solution at the first two levels looks like:
\begin{align}
A_{1,\l+1},A_{1,\l-1} =&\frac{(\DD+\l)(\l+2\nu)}{4(\l+\nu)}, \frac{(\DD-\l-2\nu)\l}{4(\l+\nu)}\ , \nn\\
A_{2,\l+2},A_{2,\l},A_{2,\l-2} = &\frac{(\DD+\l)(\DD+\l+2)^2 (\l+2\nu)(\l+2\nu+1)}{32(\DD+\l+1)(\l+\nu)(\l+\nu+1)},\nn\\
& \frac{(\DD+\l)(\DD-\l-2\nu)[(\DD-\nu)\l(\l+2\nu) + (\DD-2\nu)(\nu-1)]}{16(\DD-\nu)(\l+\nu+1)(\l+\nu-1)},\nn\\
&\frac{(\DD-\l-2\nu)(\DD-\l-2\nu+2)^2\l(\l-1)}{32(\DD-\l-2\nu+1)(\l+\nu)(\l+\nu-1)}.
\label{eq:n12}
\end{align}
Notice that low spins do not require a separate treatment: the coefficients which ``do not exist", like $A_{1,l-1}$ for $l=0$ and $A_{2,l-2}$ for $l=0,1$ come out automatically zero. This follows from the fact that $\gamma^-_{E,0}=0$ and so \reef{eq:D1act} makes sense also for $j=0$.

The recursion \reef{eq:recursion} has been found previously by Dolan and Osborn \cite{DO2}, Eq.~(3.12), who arrived at the ansatz \reef{eq:ansatz} as the way to diagonalize the homogeneous part of $\calD$. They were expanding in Jack polynomials symmetric functions in two variables $z, \bar z$, which are identical to our $\mca{P}_{E,j}$. They also give a closed-form solution of this recursion, Eq.~(3.19), which is however rather complicated (it involves ${}_4F_{3}$). In practice, it may be faster to evaluate the coefficients directly from the recursion.

\subsection[Decoupling of descendants for the leading twist]{Decoupling of descendants for the leading twist\footnote{This section is independent of the main line of reasoning and can be skipped on the first reading.}}

One interesting special case where the recursion can be solved easily is for the ``leading twist" operators $\calO$ of dimension
\beq
\Delta=\l + d-2,\qquad \l=0,1,2\ldots
\label{eq:mintwist}
\eeq
In this case we find that at each level, only the maximal allowed spin $j=\l+n$ has a nonzero coefficient. At the first two levels, this can be seen happening in Eq.~\reef{eq:n12}. For general $n$, this single nonzero coefficients takes the form:\footnote{For $d=3$, this result is agreement with the integral representation in \cite{DO3}, Eq.~(6.20).}
\beq
A_{n,\l+n}=\frac{(\l+\nu)_n (\l+2\nu)_n}{n!(2\l+2\nu)_n}\,.
\eeq

The massive decoupling of descendants implied by this result can be understood as follows. The descendants at level $n$ are obtained by acting with $n$ derivatives
\beq
\del_{\mu_1}\del_{\mu_2}\ldots\del_{\mu_n}\calO\,.
\eeq
If a $\mu_i$ is contracted with an index of $\calO$, such a state simply vanishes, because for $\l\ge1$ the dimension \reef{eq:mintwist} is the minimal value allowed by the unitarity bound and corresponds to a conserved current. 
If some of the $\mu_i$ are contracted with each other, we get a state involving $\del^2$ which has spin strictly less than $\l+n$. We should show that such states decouple. Since they do not have zero norm, this can only happen via vanishing of the matrix elements in \reef{eq:cylmat1}. Equivalently, this means that the following limit of the three point function should vanish:
\beq
\lim_{x_1\to\infty} |x_1|^{2\Delta_{\phi}}\langle \phi(x_1) \phi(x_2) \del_y^2 \calO(y)\rangle=0\,.
\label{eq:limit}
\eeq
Since the three point function $\langle \phi\phi\calO\rangle$ is known explicitly (see e.g.~\cite{DO1}), this is easy to check. Sending $x_1\to\infty$, $x_2\to0$, the three point function becomes
\begin{align}
\langle \phi|\phi(0) \calO_{\mu_1\ldots\mu_\l}(y)\rangle&=\lambda_\calO
({y_{\mu_1} \dotsm y_{\mu_{\l}}}/{|y|^{d-2+2\l}}  - \text{traces})\,\nn\\
&\propto \pd_{\mu_1}\! \dotsm \pd_{\mu_\l} \frac{1}{|y|^{d-2}}\,.
\end{align}
That the second line takes care of the trace subtractions in the first line (up to a constant factor) is obvious: it gives a tensor which has the right scaling in $y$ and is also automatically traceless (as well as conserved), due to the fact that the function 
${1}/{|y|^{d-2}}$ is harmonic in $d$ dimensions. For the same reason, this formula implies that $\del^2$-descendants decouple.

We should stress that the decoupling of $\del^2$-descendants at leading twist is peculiar to the kinematic configuration of Fig.~\ref{fig:z}. In particular, it will not happen when the points are inserted symmetrically with respect to the origin, as in the next section. This is because Eq.~\reef{eq:limit} is only true in the infinite $x_1$-limit.
  
\section{Conformal blocks in the $\rho$ coordinates}
\label{sec:rho}
\begin{figure}[htbp]
\begin{center}
\includegraphics[scale=1]{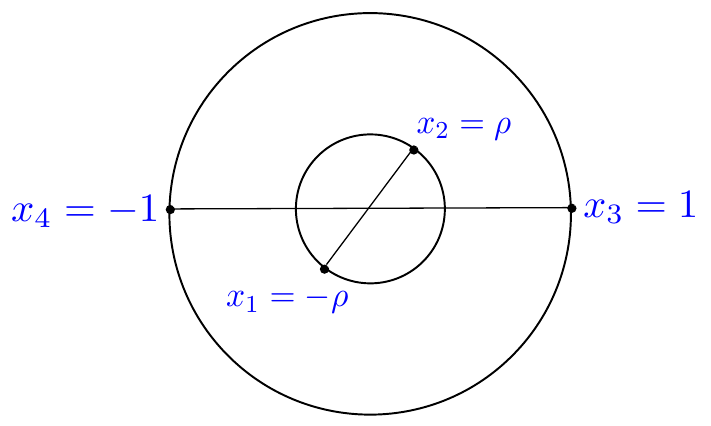}
\caption{This more symmetric configuration of operation insertions can be obtained from the one in Fig.~\ref{fig:z} by a global conformal transformation.}
\label{fig:rho}
\end{center}
\end{figure}

We now wish to analyze the four point function \reef{eq:fourpt} in a different, more symmetric, configuration of operator insertions, shown in Fig.~\ref{fig:rho}. Applying a conformal transformation, the configuration of Fig.~\ref{fig:z} can be mapped to the new one. There is a one-to-one correspondence between the complex parameters $z$ and $\rho$, fixed by demanding that the cross ratios should agree. We find 
\beq
\rho=\frac{z}{(1+\sqrt{1-z})^2}\quad\Leftrightarrow\quad z=\frac{4\rho}{(1+\rho)^2}\,.
\label{eq:rhoz}
\eeq

\begin{figure}[htbp]
\begin{center}
\includegraphics[scale=1]{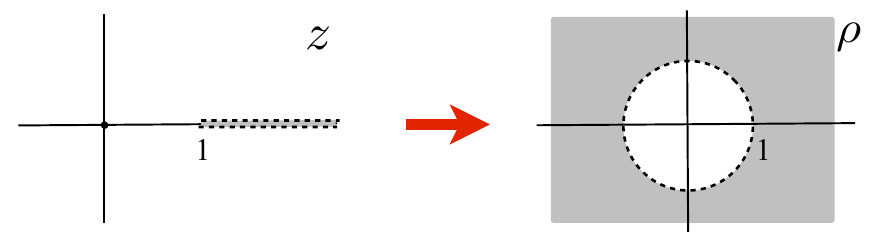}
\caption{The $\rho$ coordinate maps the regularity domain $X$ onto the unit disk.}
\label{fig:zrho}
\end{center}
\end{figure}

The $\rho$ coordinate was considered in \cite{Pappadopulo:2012jk}, where it was used to give an optimal estimate for the convergence rate of the decomposition of a four point function as a sum of conformal blocks, Eq.~\reef{eq:g_sum}. Here we will use $\rho$ to analyze the blocks themselves. As discussed at the end of section \ref{sec:zgen}, the blocks are expected to be regular in the region $X=\mathbb{C}\backslash(1,+\infty)$. The function $\rho(z)$ maps this region onto the unit disk (see Fig.~\ref{fig:zrho}). This suggests that this coordinate should be particularly suitable to analyze the blocks. To begin with, conformal block representations as power series in $\rho$ will converge for $|\rho|<1$, which is the full region of interest. Other advantages will be discussed below.

\begin{figure}[htbp]
\begin{center}
\includegraphics[scale=1]{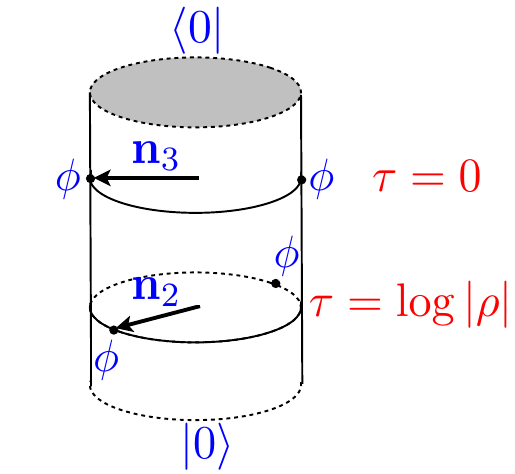}
\caption{The analogue of Fig.~\ref{fig:radz} for the new configuration.}
\label{fig:rhocyl}
\end{center}
\end{figure}

Fig.~\ref{fig:rhocyl} shows what the configuration of Fig.~\ref{fig:rho} looks like after the Weyl transformation to the cylinder. 
This picture is similar to Fig.~\ref{fig:radz} in that both the initial and final state are characterized by just one unit vector.
For this reason the whole discussion of section \ref{sec:zgen} expressing the exchanges of spin $j$ states in terms of Gegenbauer polynomials goes through unchanged. We can therefore state the following analogue of Eqs.~\reef{eq:Gexp},\reef{eq:ansatz}: conformal block of a dimension $\Delta$, spin $\l$ primary will have an expansion:
\beq
G_{\Delta,l}=\sum_{n=0}^\infty \sum_j B_{n,j} \calP_{\Delta+n,j}(r,\eta),\qquad B_{n,j}\ge0,
\label{eq:ansnew}
\eeq
where
\beq
\qquad r\equiv|\rho|,\quad\eta=\cos\arg \rho\,.
\eeq
The non-negative coefficients $B_{n,j}$ in this new expansion will of course be different from $A_{n,j}$. The spins $j$ at level $n$ will still be subject to the constraint \reef{eq:j}. However, notice that only even spin states can be exchanged since the initial state is symmetric with respect to $\rho\to-\rho$.\footnote{In fact the exchange $1\leftrightarrow 2$ corresponds to $z\to z/(z-1)$, which is equivalent to $\rho\to -\rho$.} We conclude that only even levels $n$ will have nonzero $B_{n,j}$.\footnote{In an analogous expansion for a four point function of non-identical primaries, states of all levels will be exchanged. However, if $\Delta_1=\Delta_2$ and $\DD_3 = \DD_4$, then again only even levels will appear. This is even though the exchanged primary may have both even and odd spin in this case.} 
This is unlike in \reef{eq:ansatz} where all levels have $A_{n,j}\ne0$. 

We now turn to the problem of determining the coefficients $B_{n,j}$. 
The first method is to convert from the old expansion \reef{eq:ansatz} whose coefficients $A_{n,j}$ we already know how to compute. From \reef{eq:rhoz}, the relevant variables are related by:
\beq
s=\frac{4r}{1+2r\eta+r^2}\,,\quad \xi=\frac{\eta(1+r^2)+2r}{1+2r\eta+r^2}\,.
\label{eq:sr}
\eeq
Substituting into \reef{eq:ansatz} and expanding the denominators, we will get a power series of the form
\beq
\sum_{n=0}^\infty r^{\Delta+n} Q_n(\eta),
\eeq
with $Q_n(\eta)$ certain polynomials in $\eta$. To extract $B_{n,j}$, we have to reexpand $Q_n(\eta)$ into the basis of Gegenbauers. This will give $B_{n,j}$ at level $n$ as a linear combination of $A_{n',j'}$ for $n'\le n$.

The second method is to set up an independent recursive procedure for $B_{n,j}$ based on the Casimir equation. The operator $\calD$ in $r,\eta$ coordinates takes the form:
\beq
\calD=\calD_0+\widetilde \calD, 
\label{eq:DDt}
\eeq
where the homogeneity-preserving part $\calD_0$ is the same as in \reef{eq:D0D1} with $s\to r$, $\xi\to \eta$. The homogeneity-increasing part is given by
\beq
 \widetilde\calD= 4 r^2\left\{ \left[\frac{1-2\eta^2+r^2}{{1+r^4-2r^2(2\eta^2-1)}} -\frac{\nu}{1-r^2}\right]r\pd_r+\frac{2\eta(1-\eta^2)}{1+r^4-2r^2(2\eta^2-1)} \pd_\eta\right\}.
\label{eq:Dtilde}
\eeq
Its action in the $\mca{P}_{E,j}$ basis will look like
\beq
\widetilde \calD \mca{P}_{E,j} = -\sum_{n=2,4,\ldots}\sum_{j'}\Gamma_{E,j}^{E+n,j'} \mca{P}_{E+n,j'}\,. 
\label{eq:Gammadef}
\eeq
The series is over positive even $n$, since only such powers of $r$ occur in the expansion of $\widetilde \calD$.

The dependence of the $\Gamma$ coefficients on $j'$ is found with the help of the following identities involving the Gegenbauer polynomials (the radial dependence of $\calP_{E,j}$ is not important here):
\begin{align}
(2\eta^2 -1) \mca{P}_{E,j} &= a^{-}_j \mca{P}_{E,j-2} + a^0_j \mca{P}_{E,j}  + a^{+}_j \mca{P}_{E,j+2}\,,\nn\\[3pt]
2\eta (1-\eta^2) \pd_\eta \mca{P}_{E,j}&=   b^{-}_j \mca{P}_{E,j-2} + b^0_j \mca{P}_{E,j} + b^{+}_j \mca{P}_{E,j+2}\,,
\label{eq:idgeg}
\end{align}
where
\begin{align}
\label{eq:recursioncoefs}
a^{-}_j &= \frac{j(j-1)}{2(j+\nu)(j+\nu-1)}, \quad a^0_j = \frac{\nu(1-\nu)}{(j+\nu+1)(j+\nu-1)}, \quad a^{+}_j = \frac{(j+2\nu+1)(j+2\nu)}{2(j+\nu+1)(j+\nu)}\,,\nn\\
b^{-}_{j} &= \frac{j(j-1)(j+2\nu)}{2(j+\nu)(j+\nu-1)}, \quad b^0_j = \frac{j(j+2\nu)\nu}{(j+\nu+1)(j+\nu-1)}, \quad b^{+}_j = -\frac{(j+2\nu+1)(j+2\nu)j}{2(j+\nu+1)(j+\nu)}.
\end{align}
For example, for $n=2$ we get
\beq 
\Gamma_{E,j}^{E+2,j-2} =  4 (E a_j^{-} -  b_j^{-}), \quad \Gamma_{E,j}^{E+2,j} = 4\left[ E(a^0_j + \nu) - b_j^0\right], \quad \Gamma_{E,j}^{E+2,j+2} = 4 (E a^{+}_j - b^{+}_j). 
\label{eq:G2}
\eeq

The recursion relation for the $B_{n,j}$ takes the form:
\beq 
\label{eq:Brecursion}
\left(C_{\DD+n,j} - C_{\DD,\l}\right) B_{n,j} = \sum_{n'=0,2,\ldots n-2}\sum_{j'} \Gamma^{\DD+n,j}_{\DD+n',j'} B_{n',j'}\,,
\eeq
At level $0$ we have the initial condition:
\beq
B_{0,j} = k\, \delta_{j \l}\,.
\eeq
We will set $k=1$, keeping in mind that the normalization of section \ref{sec:z} would correspond to $k=4^\Delta$.

To find the $B_{n,j}$ up to level $N$, one needs first to compute the coefficients $\Gamma_{E,j}^{E+n,j'}$ for $n\le N$. 
For example, Eq.~\reef{eq:G2} is sufficient to find the solution for level 2:
\begin{align}
B_{2,\l-2} &= \frac{\l(\l-1)(\DD-\l-2\nu)}{2(\l+\nu-1)(\l+\nu)(\DD-\l-2\nu+1)}, \qquad 
B_{2,\l} = \nu\frac{\DD\nu(\nu-1) +(\DD-1)\l(\l+2\nu)}{(\DD-\nu)(\l+\nu+1)(\l+\nu-1)}, \nonumber\\
B_{2,\l+2} &= \frac{(\DD+\l)(\l+2\nu)(\l+2\nu+1)}{2(\DD+\l+1)(\l+\nu)(\l+\nu+1)}.
\label{eq:Bs}
\end{align}

\subsection{Comparison between the $z$ and $\rho$ expansions}

We have presented two ways to expand the conformal blocks: the ``$z$-series" \reef{eq:ansatz} and the ``$\rho$-series" \reef{eq:ansnew}. We will now argue that the second expansion is more efficient, in the sense that it converges more rapidly and fewer terms need to be evaluated in order to get a good approximation. This happens because of the better choice of the expansion parameter and the better asymptotic behavior of the series coefficients.

Let us start with the expansion parameters. The interesting range for the $\rho$ coordinate is the unit disk $|\rho|<1$.
The $\rho$-series will converge absolutely, everywhere in this disk. For any $\eps>0$ the convergence will be uniform for $|\rho|<1-\eps$. To prove this statement, consider first $\rho=1-\eps$ real. For such $\rho$ all terms in the series are positive, and so the series must converge: a divergence here would mean a physical singularity for the conformal block, while as we discussed all such singularities are confined to $|\rho|=1$. Convergence everywhere else in the disk will be only better. 
Physically, this follows by the Cauchy inequality: every term in the series is the product of two matrix elements, which for real $\rho$ become Hermitian conjugates of each other. Formally, this is because the Gegenbauer polynomials with our normalization are less than one in the absolute value on the interval $[-1,1]$. 

The same argument can be used to show that the $z$-series will converge absolutely in the disk $|z|<1$. As we discussed, this does not even cover the full regularity region of the conformal blocks. Moreover, from the second Eq.~\reef{eq:rhoz} we have
\beq
|z(\rho)/\rho|>1 \qquad(|\rho|<1)\,.
\eeq
So even in the region where both series converge, the $\rho$-series will always have a strictly smaller expansion parameter.
 
An additional bonus appears when considering conformal blocks for equal external dimensions. As we have seen, in this case the $\rho$-series involves only even levels. So, the effective expansion parameter becomes $\rho^2$. In conformal bootstrap applications, one usually uses conformal blocks evaluated near $z=1/2$, which would correspond to $\rho=3-2\sqrt{2}\approx 0.17$ and $\rho^2\approx 0.03$.

Let us now examine the expansion coefficients. We are interested in their asymptotic behavior when $\Delta$ or $l$ become large. In the large $\Delta$ limit the coefficients $A_{n,j}$ at level $n$ grow as
\beq
A_{n,j}=O(\Delta^n)\,.
\label{eq:asA}
\eeq
For $n=1,2$ this can be seen in Eqs.~\reef{eq:n12}. The reason for this growth is that the operator $\calD_1$ is second order in $\del_s$. Because of this the coefficients $\gamma_{E,j}^\pm$ in \reef{eq:D1act}
are $O(E^2)$. On the other hand the factor in the RHS of the recursion relation:
\beq
C_{\DD+n,j} - C_{\DD,\l} =2n \Delta + n(n-d) +j(j+d-2)-l(l+d-2)
\label{eq:RHSc}
\eeq
increases only linearly in $\Delta$. So, going up one level in $n$, the coefficients $A_{n,j}$ gain one power in $\Delta$.

Turning to the second expansion, we encounter a crucial difference. Unlike $\calD_1$, the operator $\widetilde{\calD}$ in \reef{eq:DDt} is only first order in $\del_r$. So the coefficients $\Gamma$ entering the $B_{n,j}$ recursion grow only linearly in $E$, and this growth cancels when dividing by \reef{eq:RHSc}. Contrary to the previous case, going one level up in the recursion relation does not increase the leading power of $\Delta$. We conclude that the coefficients $B_{n,j}$ remain bounded in the large $\Delta$ limit. For $n=2$ this is illustrated by Eqs.~\reef{eq:Bs}.

Keeping more careful track of the size of the relevant factors, one can show the following sharper statement (see appendix \ref{sec:largeL}). Each coefficient $B_{n,j}$ is uniformly bounded in the full range of $\Delta$ and $l$ allowed by the unitarity bounds, with the bound depending only on the level $n$ and on $d$:
\beq
\max_{\Delta,l} \Bigl(\max_j B_{n,j}\Bigr)\le b(n,d)\,.
\label{eq:bounded}
\eeq
The region close to the free scalar limit $l=0$, $\Delta\to\nu$ is understood excluded when taking the maximum. As is well known, the scalar conformal block becomes singular in this limit. Physically this is due to the fact that the free scalar must be decoupled from everything else. In our representation, the singularity first shows up in the coefficient $B_{2,0}\sim(\Delta-\nu)^{-1}$, see Eq.~\reef{eq:Bs}, and then feeds into higher levels.

Let us discuss a bit how the coefficients $B_{n,j}$ grow with $n$. This growth is related with the the behavior of the conformal block for real $\rho\to 1$. It can be shown using the results of Ref.~\cite{ElShowk:2012ht}\footnote{This follows for $l=0$ from the explicit ${}_3F_2$ representation on the real line, Eq.~(4.10) of \cite{ElShowk:2012ht}, and remains valid for $l\ge1$ by the recursions in Appendix A of \cite{ElShowk:2012ht}.} that in this limit the conformal block has a power-like singularity of the form:
\beq
G_{\Delta,l}(\text{real }\rho\to 1)\sim \frac1{(1-\rho)^{d-2}}\,\qquad\left(\log \frac1{1-\rho}\text{ for }d=2\right).
\eeq
On the other hand, the $\rho$-series representation for real $\rho>0$ takes the form:
\beq
G_{\Delta,l}(\text{real }\rho>0)=\sum_{n=0}^\infty \beta_n \rho^{\Delta+n},\qquad \beta_n=\sum_j B_{n,j}\,.
\eeq
From compatibility with the $\rho\to 1$ asymptotics, we conclude that for any $\Delta$ and $l$ the sum of the coefficients at level $n$ behaves asymptotically as
\beq
\beta_n\sim n^{d-3}\qquad(n\to\infty)\,.
\label{eq:betan}
\eeq
It would be interesting to know how the ratio $\beta_n/n^{d-3}$ behaves for small and intermediate $n$.
The simplest possibility which accommodates both \reef{eq:bounded} and \reef{eq:betan} is that $\beta_n\le c(d) n^{d-3}$ for all $n$, $\Delta$ and $l$. However, further study is needed to check this hypothesis.

To finish this section, we would like to demonstrate how the highlighted differences between the $z$- and $\rho$-series can be seen in the explicit expressions for the conformal blocks available for even $d$. These expressions  \cite{Ferrara:1974ny,DO1,DO2,DO3} are written in terms of the functions 
\beq
k_a(z)=z^{a/2}{}_{2}F_{1}(a/2,a/2;a;z)
\label{eq:ka}
\eeq
with $a=\Delta+l$ and $\Delta-l-2\nu$. For large $\Delta$, the $z^n$ coefficient in the expansion of the ${}_2 F_{1}$ grows as $a^n\sim \Delta^n$. This is the same growth as in \reef{eq:asA}. However, when the $\rho$ variable is used, the function $k_a$ can be transformed using a hypergeometric identity
\beq
k_a[{4\rho}/{(1+\rho)^2}]= (4\rho)^{a/2}
{}_{2}F_{1}(1/2,a/2;(a+1)/2;\rho^2)\,.
\eeq
As advertised, this is a function of $\rho^2$ and the expansion coefficients do not grow with $a$.

\subsection[Relation to Zamolodchikov's uniformizing variable]{Relation to Zamolodchikov's uniformizing variable\footnote{This section is independent of the main line of reasoning and can be skipped on the first reading.}}
\label{sec:Zam}

We would like to briefly mention a similarity between the change of variables from $z$ to $\rho$ advocated here,
and the one proposed long ago by Al.~Zamolodchikov \cite{Zamolodchikov:1987} in the study of 2$d$ ``big" conformal blocks. His variable is given by
\beq
q=e^{i\pi\tau},\quad \tau=i K(1-z)/K(z)\,,
\eeq
where $K(z)$ is a complete elliptic integral of the first kind
\beq
K(z)=\frac 12\int_0^1 \frac{dt}{[t(1-t)(1-z t)]^{1/2}}\,.
\eeq
The variable $\tau$ takes values of the upper half plane, parametrizing the universal covering of the Riemann sphere with three punctures 0,1, and $\infty$. The point is that in the 2$d$ case, conformal blocks factorize as $\calF(z)\calF(\bar z)$ where $\calF(z)$ is holomorphic in the complex plane with branch points at the punctures. So it is natural to view it as an analytic function on the universal covering space. The conformal blocks are then given as power series in $q$. Since $|q|<1$ in the upper half plane, these series converge everywhere where $\calF(z)$ is analytic, while power series representations in $z$ converge only for $|z|<1$.

\begin{figure}[htbp]
\begin{center}
\includegraphics[scale=0.7]{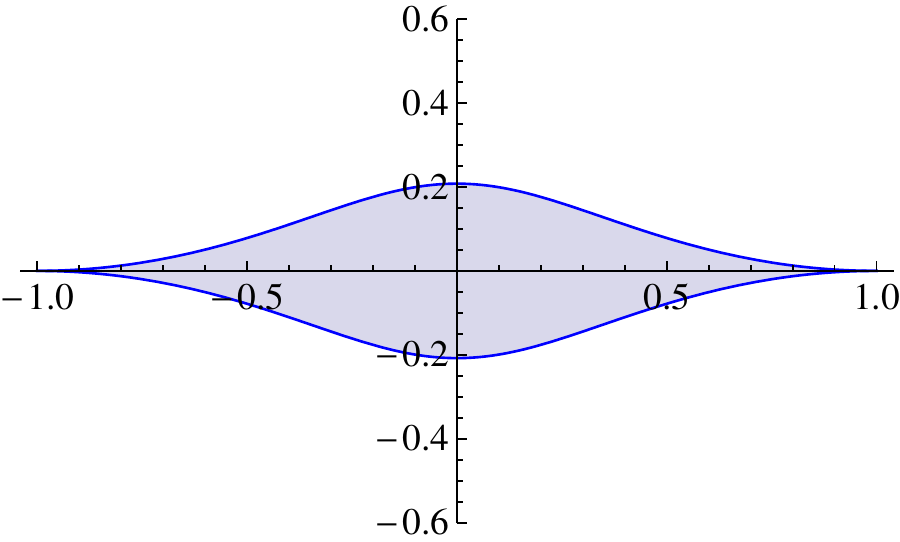}
\caption{The regularity domain $X$ is mapped in the $q$ plane onto this UFO-shaped region inside the unit disk. For example, the point $z=1/2$ is mapped to $q\approx 0.043$.}
\label{fig:q}
\end{center}
\end{figure}

In the 2$d$ case, the variable $q$ is a more efficient expansion parameter than our variable $\rho$. For example, the conformal block regularity domain $X=\mathbb{C}\backslash (1,+\infty)$ is mapped on a subset of the complex plane located strictly inside the unit disk, see Fig.~\ref{fig:q}. For general $d$, when holomorphic structure is absent, the variable $\rho$ is probably best possible. 

\section{Potential applications to the conformal bootstrap}
\label{sec:appl}
In the previous section, we introduced a new way to represent the conformal blocks, by expanding them in the polar coordinates associated with the complex variable $\rho$. Our interests in the blocks stems from the role they play in the conformal bootstrap program. We believe that our new representation will turn out quite useful in this context. Here we will list several ideas, leaving their complete development for the future.

Most existing applications of the conformal bootstrap to CFT in $d\ge3$ dimensions followed the following scheme proposed in \cite{Rattazzi:2008pe}. Due to the complexities of dealing with the full bootstrap system, one focuses on just one equation out of the infinitely many pictured in Fig.~\ref{fig:ass}: the one in which all four external states are one and the same, scalar primary $\phi$.\footnote{The case when the external states are different components of a scalar global symmetry multiplet has also been considered \cite{Poland:2010wg,Rattazzi:2010yc,Vichi:2011ux,Poland:2011ey}.} This equation is obtained by substituting the conformal block expansion \reef{eq:g_sum} into the crossing symmetry constraint \reef{eq:cross} and takes the form:
\beq
(v^{\Delta_\phi}-u^{\Delta_\phi})+ \sum_{i} f_{{i}}^2 [v^{\Delta_\phi} G_{\Delta_i,l_i}(u,v)-(u\leftrightarrow v)]=0\,.
\label{eq:sumrule}
\eeq
The sum is over all primary operators $\calO_i$ appearing in the OPE $\phi\times\phi$, with $\Delta_i,l_i,f_i$ their dimensions, spins, and OPE coefficients. The unit operator contribution is separated explicitly.

In the approach of \cite{Rattazzi:2008pe}, one views \reef{eq:sumrule} as an equation for infinitely many unknowns $f_i^2\ge 0$. 
For any given spectrum of $\calO_i$'s one can ask if a solution exists, and if not, then a CFT with such a spectrum is impossible. It turns out that even mild assumptions, like the absence of a scalar primary below a certain dimension, can lead to an inconsistent spectrum. 
Making one OPE coefficient too large can also produce an inconsistency. It would take too much time to review in detail the results of these studies (see \cite{Rattazzi:2008pe,Rychkov:2009ij,Caracciolo:2009bx,Poland:2010wg,Rattazzi:2010gj,
Rattazzi:2010yc,Vichi:2011ux,Poland:2011ey,ElShowk:2012ht,ElShowk:2012hu}), but we would like to highlight here two basic issues 
which were important in all of them.

First, the functional equation \reef{eq:sumrule} looks still rather complicated, and in practice one has to replace it by a finite-dimensional constraint, which will of course be weaker but hopefully more tractable. One simple way to do it would be to choose a large finite number of points in the $(u,v)$ plane and impose \reef{eq:sumrule} at each of these points. Ref.~\cite{Rattazzi:2008pe} followed another way, which became standard: to impose \reef{eq:sumrule} in Taylor expansion up to a fixed large order around just one, well-chosen point. The obvious choice for such a point is $z=\bar z=1/2$, i.e.~$u=v=1/4$, which is invariant under the crossing symmetry transformation $z\to1-z$.

The second issue is that the conformal blocks appearing in \reef{eq:sumrule} are complicated functions, even for $d=4$ where explicit expressions in terms of hypergeometrics are known. Some simplifications occur for specific values of $\Delta$, but this does not help, since the dimensions of $\calO_i$ are unknown and should be allowed to vary freely between the unitarity bound and infinity. For generic $\Delta$, conformal block derivatives at $z=\bar z=1/2$ must be evaluated numerically.\footnote{For even $d$, these derivatives can be written via the ${}_3F_2$ functions (see appendix B.1 of \cite{Poland:2010wg}). It is not known at present how to use these analytic expressions in practice, rather than as a starting point for the numerical evaluation.} For this reason all the studies cited above used numerical analysis.\footnote{We should also mention three different types of bootstrap analyses where analytic results could be obtained: \cite{Heemskerk:2009pn} in the large $N$ expansion; \cite{Liendo:2012hy} for CFT in presence of a boundary; \cite{Fitzpatrick:2012yx} and \cite{Komargodski:2012ek} in the Minkowski space near the light cone.} Moreover, this evaluation is an expensive operation and often presents a computational bottleneck.

We will now describe new ways of approaching these issues, made possible by the $\rho$-series representation.

\subsection{Inexpensive derivative evaluation for all $\Delta$ and $l$}

Let us briefly review the existing ways of evaluating conformal blocks and their derivatives at the point $z=\bar z=1/2$. For even $d$, one uses the explicit representations of Dolan and Osborn \cite{DO1,DO2}. For $d=4$ they take the form:
\beq
G_{\Delta,l}(z,\bar z)= \frac{z\bar{z}}{z-\bar z}[k_{\Delta+l}(z) k_{\Delta-l-2}(\bar{z})-(z\leftrightarrow \bar{z})]\,.
\label{eq:DO}
\eeq
By this formula, partial derivatives of $G_{\Delta,l}(z,\bar z)$ can be represented as quadratic forms in the derivatives of the function $k_a(z)$, defined in Eq.~\reef{eq:ka}. One can now create an interpolated lookup table of $k_a(z)$'s derivatives at $z=1/2$ for a range of $a$. This is a time-consuming operation, because the hypergeometric function in \reef{eq:ka} is expensive to evaluate. However, one needs to do this only once. Once the table is created and stored, partial derivatives of \reef{eq:DO} can be computed quickly for any $\Delta$ and $l$. Such a strategy was used in \cite{Rychkov:2009ij,Caracciolo:2009bx,Rattazzi:2010gj,Rattazzi:2010yc,Vichi:2011ux}, and a similar one in \cite{Poland:2010wg}. More recently, Ref.~\cite{Poland:2011ey} found a way to dispense with the lookup table altogether, computing the derivatives of $k_a(z)$ at $z=1/2$
via a very rapidly convergent infinite product representation. 
 
Turning to general $d$, a method to evaluate conformal block derivatives was developed last year in \cite{ElShowk:2012ht}, where it 
 was used to study the 3$d$ Ising model. This method combines a variety of ideas. One begins by evaluating partial derivatives along the $z=\bar z$ line, first for $l=0$ and $l=1$ using explicit ${}_3 F_2$ expressions found by \cite{ElShowk:2012ht}, then for higher $l$ using the recursion relations from \cite{DO3} reducing those blocks to the lower-spin ones. Then, partial derivatives in the orthogonal direction are computed \textit{\`a la} Cauchy-Kovalevskaya, using the fact that the conformal blocks satisfy a second-order partial differential equation. 
 
The $\rho$-series gives a new way to evaluate conformal blocks and their derivatives, which works for general $d$ and around any $z$. To achieve the necessary precision, one needs to evaluate the coefficients $B_{n,j}$ as a function of $\Delta$ and $l$ up to a sufficiently high order, using the recursion relation \reef{eq:Brecursion}. It is important that the necessary number of terms will be independent of $\Delta$ and $l$, because of the bound \reef{eq:bounded}. For example, to be able to compute the conformal blocks with double precision ($10^{-16}$) one would need the coefficients up to level
\beq
n\approx 16/\log_{10}(1/\rho)\,,
\eeq
which gives $n\approx20$ for $z=1/2$. This number  is a bit of an underestimate, because it assumes that the sum of the coefficients at level $n$ is uniformly bounded, while in fact it grows with $n$ as in Eq.~\reef{eq:betan}. Also more levels will be needed if one wants to evaluate derivatives.

It should also be rather easy to generalize the $\rho$-series method to the case of unequal external dimensions $\Delta_1\ne \Delta_2$ and $\Delta_3\ne \Delta_4$. The extra terms in the Casimir operator for unequal external dimensions are all first order in derivatives \cite{DO2}. So, the operator $\widetilde\calD$ will remain first order, and we can expect that the boundedness properties of the coefficients $B_{n,j}$ will still hold. Such a generalization will be useful for the conformal bootstrap analysis of several scalar correlators simultaneously. For example, in the 3$d$ Ising model, it would be interesting to study simultaneously the correlators $\langle \sigma\sigma\sigma\sigma\rangle$, $\langle \sigma\sigma\eps\eps\rangle$ and $\langle \eps\eps\eps\eps\rangle$. The second of these correlators has unequal external dimensions in two conformal partial wave expansion channels out of three. 

\subsection{Analytic toy model for the conformal bootstrap}
\label{sec:toy}
As already mentioned, Eq.~\reef{eq:sumrule} is usually analyzed numerically. Numerics are essential both for computing the conformal blocks and their derivatives at $z=1/2$, and for performing searches in the resulting derivative spaces, which in the most advanced studies \cite{Poland:2011ey,ElShowk:2012ht} can have $O(100)$ dimensions.  If one needs precision, numerics are unavoidable at present.

Leaving precision aside, here we would like to address a more modest question: can one provide an analytic understanding of \textit{why} the method of \cite{Rattazzi:2008pe} gives nontrivial constraints? Some intuitive explanations were given in section 5.1 of \cite{Rattazzi:2008pe}, but those still relied on properties of conformal blocks which had to be checked by plotting them in Mathematica. As we will now explain, analytic results can be obtained using the $\rho$-series representation.

The basic idea is the following. We have shown that the conformal blocks can be represented as series in the ``Gegenbauer blocks" $\calP_{E,j}(r,\eta)$. These series are rapidly convergent, to the extent that even the first term provides already a very good approximation:
\beq
G_{\Delta,l}\approx \calP_{\Delta,l}(r,\eta)\,.
\eeq
The relative error is of order $\rho^2$, which is about $3\%$ at $z=1/2$. Importantly, the error is uniformly small for all $\Delta$ and $l$. Replacing the conformal blocks  in Eq.~\reef{eq:sumrule} by their Gegenbauer block approximation, we get the ``toy" bootstrap equation ($\delta\equiv\Delta_{\phi}$):
\begin{gather}
[(1-z)^{\delta}(1-\bar z)^{\delta}-z^{\delta}{\bar z}^{\delta}]+ \sum_{\Delta,l} f_{{\Delta,l}}^2 [H_{\Delta,l}(z,\bar z)-
H_{\Delta,l}(1-z,1-\bar z)]=0\,,\nn\\
H_{\Delta,l}(z,\bar z)\equiv (1-z)^{\delta}(1-\bar z)^{\delta} [(\rho(z)\rho(\bar z)]^{\Delta/2} 
C^{(\nu)}_l\left(\frac{\rho(z)+\rho(\bar z)}{2[\rho(z)\rho(\bar z)]^{1/2}}\right)/C^{(\nu)}_l(1)\,.
\label{eq:toyfull}
\end{gather}
 It is expected to give qualitatively the same results as the full bootstrap equation, with an advantage that the analysis can be done analytically, since the Gegenbauer block derivatives can be computed explicitly. We will give below two examples of this approach.

A comment is in order concerning the spectrum of $\Delta$'s and $l$'s appearing in the second term of the toy bootstrap equation. If we view it as an \textit{approximation} to the full equation \reef{eq:sumrule}, then of course it's the same spectrum as in the full equation, i.e.~all primaries in the $\phi\times\phi$ OPE. However, an alternative point of view can be useful. We can consider the toy equation as \textit{exact}, provided that we allow not only the primaries but also their descendants to appear in the spectrum. When we do the bootstrap in terms of the full conformal blocks, we have extra constraining power because the OPE coefficients of descendants are proportional to those of the primaries. In the toy bootstrap, we choose to discard this information and allow descendants to appear with independent coefficients. One can also imagine an intermediate situation, when descendants up to a certain level are included with the relative coefficients fixed by conformal symmetry, while the higher ones are taken independent.

\subsubsection{Toy bootstrap for $z=\bar z$}

For the first example \cite{EPFLlectures}, let us consider the toy equation \reef{eq:toyfull} for real $0<z<1$. The angular part of the Gegenbauer blocks is then trivial, and the equation takes an extremely simple form:
\beq
[(1-z)^{2\delta}-z^{2\delta}]+ \sum_{\Delta} f_{{\Delta}}^2 \{(1-z)^{2\delta} [\rho(z)]^{\Delta}-z^{2\delta} [\rho(1-z)]^{\Delta}\}=0\,.
\label{eq:toy}
\eeq
We will use this equation to show that there is an upper bound on the lowest primary dimension $\Delta_{\min}$ in the $\phi\times\phi$ OPE. This is a problem of the kind first considered in \cite{Rattazzi:2008pe}, except that here we are not distinguishing between scalar and higher spin primaries. 

For the proof, let us Taylor expand Eq.~\reef{eq:toy} in $x=z-1/2$. Only odd powers of $x$ will appear since the functions are odd. From the first term we get:
\beq
\label{expx}
(1-z)^{2\delta}-z^{2\delta}=-C_\delta\left(x+{\textstyle\frac{4}{3}}(\delta-1)(2\delta-1)x^3+\ldots\right)\,,
\eeq
where $C_\delta>0$ is a constant whose precise value is unimportant. When expanding the Gegenbauer block terms, let us assume that all $\Delta\gg\delta$ (we aim for a contradiction here). Then we can approximate $z^{2\delta}\approx (1-z)^{2\delta}\approx (1/2)^{2\delta}=const$, while the relevant part is:
\beq
\label{Dbiggethand}
[\rho(z)]^\Delta-[\rho(1-z)]^\Delta= B_\Delta \left(x+{\textstyle\frac{4}{3}}\Delta^2x^3+\ldots\right) \ ,
\eeq
where $B_\Delta>0$ is another inessential constant. We can change normalization of the blocks so that $B_\Delta\to1$ (incorporating this constant into $f_\Delta^2$). Requiring that \reef{eq:toy} be satisfied term by term in the Taylor expansion, we get:
\beq
\label{eq:t1}
C_\delta=\sum  f_{{\Delta}}^2\,,\qquad C_\delta (\delta-1)(2\delta-1)=\sum \Delta^2f_{{\Delta}}^2\,.\nn
\eeq
The $O(x^5)$ terms etc.~would give more equations but we won't use them here. Bounding the RHS of the second equation from below by $ \Delta^2_{\min} \sum  f_{{\Delta}}^2$ and using the first equation, we conclude that
\beq
\Delta_{\min}\le \sqrt{(\delta-1)(2\delta-1)}\,.
\eeq
This shows that our original assumption that $\Delta_{\min}\gg \delta$ is inconsistent, hence there must be a bound on
$\Delta_{\min}$ in terms of $\delta$. Its actual value can be found by a more careful analysis, expanding the Gegenbauer block terms without the approximation $\Delta\gg\delta$. 

\subsubsection{Including the spin dependence}

We will next consider the toy bootstrap equation \reef{eq:toyfull} not restricting to the $z=\bar z$ line. The advantage is that the spin information will be now accessible through the order of the Gegenbauer polynomial. So we can try to set a upper bound on the lowest \textit{scalar} in the $\phi\times\phi$ OPE, which is precisely the problem considered in \cite{Rattazzi:2008pe}. 

For the simplest bound, we will expand \reef{eq:toyfull} to the third order in $z$ and $\bar z$ around $z=\bar z=1/2$. Because of various (anti)symmetries, only three derivatives are independent. We will choose 
\beq
\del_z,\ \del^3_z-\del^2_z\del_{\bar z},\  \del^3_z+3\del^2_z \del_{\bar z}
\eeq 
as a basis, as these linear combinations somewhat simplify the subsequent algebra. Because all functions are elementary, the derivatives can be evaluated explicitly. Equating them to zero, we get the following linear system:
\begin{align}
&\sum_{\Delta,l}\rho(1/2)^{\Delta} f^2_{{\Delta,l}}\, \mathbf{h}_{\Delta,l}=\mathbf{h}_{0,0}\,,
\label{eq:linsys}\\
&\mathbf{h}_{0,0}=[2 \delta ,\, -16\delta( \delta -1) ,\, 16\delta(2 \delta ^2-3 \delta+1)
]^t\,,\nn\\
&\mathbf{h}_{\Delta,l} =[\sqrt{2} \Delta -2 \delta,\,
-4 \Delta ^2+\Delta  \sqrt{2}(-4  \delta +8  k_l+7)+16 \delta ^2-16\delta (k_l+1)-8
   k_l,\nn\\
   &\hspace{1cm}8 \sqrt{2} \Delta ^3-12(4 \delta +1) \Delta ^2+\sqrt{2}(48
   \delta ^2-12 \delta +7 ) \Delta -16\delta(2 \delta ^2-3 \delta+1) ]^t\nn\,,
\end{align}
where $k_l$ is the logarithmic derivative of the Gegenbauer polynomials at $\eta=1$:
\beq
k_l\equiv \frac{[C_l^{(\nu)}]'(1)}{C_l^{(\nu)}(1)}=\frac{l(l+2\nu)}{2\nu+1}\,.
\eeq
Notice that the spin and the spacetime dimension enter only through this coefficient and via the unitarity bounds.

Next we eliminate the RHS from the last two equations in \reef{eq:linsys} with the help of the first one. We get a homogeneous system:
\begin{gather}
\sum_{\Delta,l} q_{\Delta,l}\, \mathbf{g}_{\Delta,l}=(0,0)^t\,,\qquad q_{\Delta,l} \equiv \Delta^3 \rho(1/2)^{\Delta} f^2_{{\Delta,l}}\,,
\label{eq:hom}\\
\mathbf{g}_{\Delta,l} =\left(
\begin{array}{l}
1-3\sqrt{2}(\delta +1/4)\Delta^{-1 }+({4 \delta ^2+{3 \delta }/{2}-{1}/{8}}){\Delta ^{-2}}\\[3pt]
{\Delta^{-1} }
-\sqrt{2}(
   \delta +2 k_l-{1}/4){\Delta ^{-2}}+(4 \delta+2) k_l{\Delta ^{-3}}
   \end{array}\right)\,.
\end{gather}
The rest of the discussion follows closely section 5.4 of \cite{Rattazzi:2008pe}. We have to study how the direction of the vectors $\mathbf{g}_{\Delta,l}$ varies when we increase $\Delta$ from the unitarity bound to infinity. 
For $l=0$ we vary $\Delta$ from $\Delta_{0,\min}$ to infinity, where $\Delta_{0,\min}$ is the lowest scalar dimensions.
If the set of all directions stays within a cone of opening angle $<\pi$, then \reef{eq:hom} will not have a nontrivial solution, while in the opposite case it will. The difference from \cite{Rattazzi:2008pe} is that now $\mathbf{g}_{\Delta,l}$ are given explicitly, and the analysis can be carried out analytically. We will not give here the details, but we have checked that an upper bound on $\Delta_{0,\min}$ can be obtained using this method for all $2\le d\le 4$, at least for $\delta$ near the scalar unitarity bound. 

\subsection{Truncated bootstrap equation with an error estimate}

As mentioned above, one could also try to do bootstrap imposing the bootstrap equation point by point at several $z=z_i$, rather than in the Taylor expansion around $z=1/2$. We would like to discuss here the issues arising if one wants to implement this technique. Conformal block evaluation for any $z$ can be done with the $\rho$-series. The next question is then how to distribute the sampling points. To get an idea, let us consider the rate of convergence of the conformal block decomposition \reef{eq:g_sum}. As shown in Ref.~\cite{Pappadopulo:2012jk}, the error induced by truncating \reef{eq:g_sum} at some maximal dimension $\Delta=\Delta_*$ is exponentially small:
\beq
\Biggl|\sum_{\calO:\Delta(\calO)\ge\Delta_*} f_{\calO}^2 G_{\calO}(z,\bar z)\Biggr|\lesssim \frac{\Delta_*^{4\Delta_\phi}}{\Gamma(4\Delta_\phi+1)}  |\rho(z)|^{\Delta_*}\,.
\label{eq:bnd}
\eeq
To be precise, this estimate was shown to hold for $\Delta_*\gg \Delta_\phi/(1-|\rho(z)|)$\,. Most importantly, it holds in an arbitrary CFT with no extra assumptions about the $\phi\times\phi$ OPE. For example, it might seem that having too many operators at high $\Delta$, or a single operator with a huge OPE coefficient, might invalidate this bound. However, the proof in \cite{Pappadopulo:2012jk} shows that such situations cannot occur in a consistent CFT.

\begin{figure}[htbp]
\begin{center}
\includegraphics[scale=0.7]{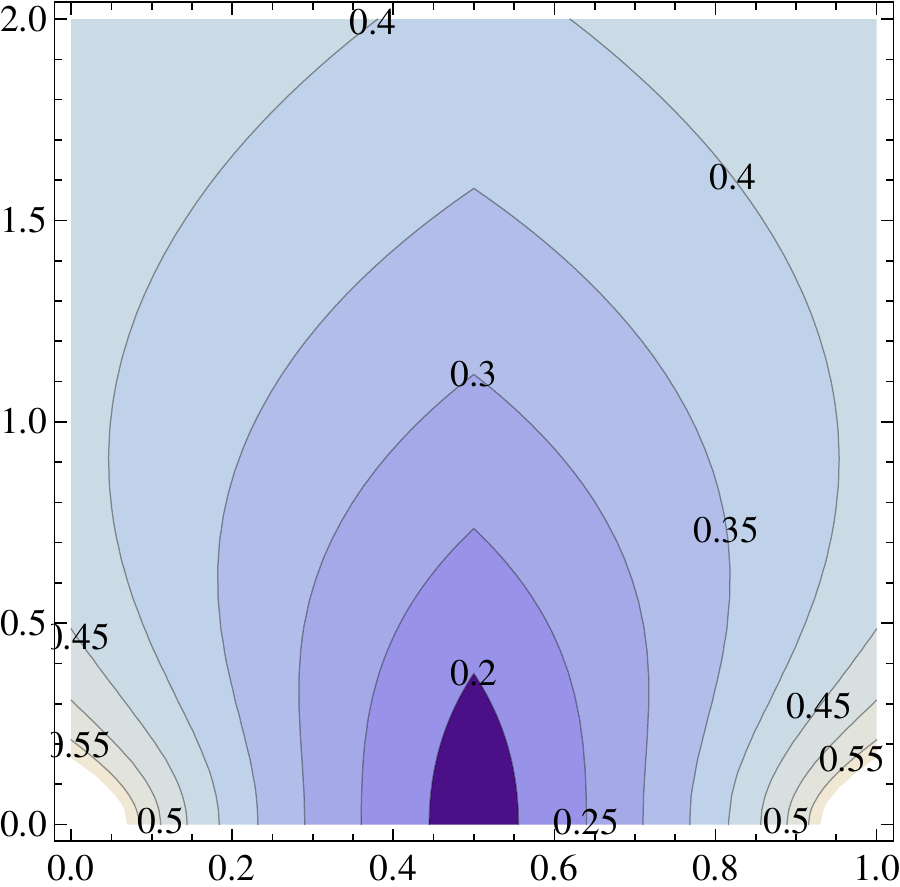}
\caption{The contour plot of the function $\lambda(z)$ in the plane $(\text{Re }z, \text{Im }z)$. Only the region $\text{Im }z\ge0$ is shown, since the conformal blocks are symmetric in $z, \bar z$. }
\label{fig:contour}
\end{center}
\end{figure}

The estimate \reef{eq:bnd} is relevant to our discussion, because in most practical approaches to the bootstrap one has to truncate the spectrum of considered operators from above (to make the problem finite). Now we know that the error induced by this operation is controlled by $|\rho(z)|$, while the error in the crossed channel will be controlled by $|\rho(1-z)|$. Therefore it seems natural to distribute the points $z_i$ in a region of the form (see Fig.~\ref{fig:contour})
\beq
\lambda(z)=\max(|\rho(z)|,|\rho(1-z)|)\le \lambda_c\,,\,
\label{eq:reg}
\eeq
where $\lambda_c$ should be chosen commensurately with the eventual dimension cutoff $\Delta_*$.

One way to choose $\Delta_*$ is so that the error \reef{eq:bnd} is below the numerical precision one is working with (say double precision) everywhere within the region \reef{eq:reg}. Alternatively, one can choose $\Delta_*$ lower, so that the error is non negligible. Then one has to include this error estimate directly into the bootstrap equation. Such a modified equation takes the form:
\begin{gather}
\Bigl|(v^{\Delta_\phi}-u^{\Delta_\phi})+ \sum_{\Delta(\calO)\le\Delta_*} f_{\calO}^2 [v^{\Delta_\phi} G_{\calO}(u,v)-(u\leftrightarrow v)]\Bigr|\le E(z,\bar z)\,,\nn\\
E(z,\bar z)\simeq \frac{\Delta_*^{4\Delta_\phi}}{\Gamma(4\Delta_\phi+1)} \max\bigl( 
|1-z|^{2\Delta_\phi}|\rho(z)|^{\Delta_*}, |z|^{2\Delta_\phi}|\rho(1-z)|^{\Delta_*}\bigr)\,.
\label{eq:trunc}
\end{gather}

We think it would be interesting to try to carry out bootstrap analysis based on this ``truncated bootstrap equation" rather on the conventional technique of expanding around $z=1/2$. There are many free parameters one can play with: $\Delta_*$, $\lambda_c$, the number of points $z_i$ at which to impose \reef{eq:trunc}, and what is the optimal way to distribute them in the region \reef{eq:reg}. Once all these parameters are fixed, the problem of deciding whether \reef{eq:trunc} has a solution with $f^2_\calO\ge0$ can be solved via the linear programming algorithms.

It's worth pointing out an additional feature of Eq.~\reef{eq:trunc}, which makes it particularly useful when the conformal blocks are computed via the $\rho$-series, whose coefficients can be computed up to arbitrary order but whose closed form is unknown. Namely, it remains valid when the conformal blocks $G_{\calO}(u,v)$ are replaced by the ``truncated blocks"---the partial sums of the $\rho$-series up to the level $\Delta+n\ge \Delta_*$. This is because the error estimate \reef{eq:bnd} is in fact valid when the contributions of all states of dimension above $\Delta_*$ are included into the LHS (and not just the conformal multiplets of primaries above $\Delta_*$). It is in this stronger form that the error estimate was proved in Ref.~\cite{Pappadopulo:2012jk}.

 \section{Discussion}
\label{sec:concl}

In this paper we developed the theory of conformal blocks rooted in their physical meaning as sums of exchanges of descendant states in the radial quantization. This point of view is standard in the 2$d$ CFT literature \cite{Belavin:1984vu}, and our goal here was to demonstrate its utility in higher dimensions. 

We explained how quantum mechanics fixes the structure of conformal block in radial coordinates: it is an integer-spaced power series in $r$ with angular dependence given by Gegenbauer polynomials. The coefficient of each term is positive as a consequence of unitarity. These coefficients are easy to find using recursion relations following from the fact that the conformal blocks are eigenfunctions of the quadratic Casimir of the conformal group.

We highlighted the existing freedom in the choice of the radial coordinates. It's the same freedom as when expanding the product of two operators $\phi(x_1) \phi(x_2)$ into a sum of operators inserted in some point $x_0$, which becomes the radial quantization origin. Each choice gives a different representation of the same conformal block, and it is not a priori clear which one is more convenient. In this paper we analyzed in detail two natural choices, the end point $x_0=x_1$, and the middle point $x_0=(x_1+x_2)/2$.

The end point choice (section \ref{sec:z}) corresponds to working with the complex variable $z$ often used to represent conformal blocks, with explicit ${}_2 F_1$ representations available in even dimensions $d$. For general $d$ considered here, we expand conformal blocks in a power series in $|z|$ times Gegenbauers. The expansion coefficients $A_{n,j}$ satisfy a three-term recursion relation, also derived earlier from a different point of view by Dolan and Osborn \cite{DO2}. An unpleasant feature of these expansions is that the coefficients at level $n$ grow with the exchanged primary dimension as $\Delta^n$. For large $\Delta$ many terms need to be evaluated to get a good approximation to the conformal block.

Choosing the middle point (section \ref{sec:rho}), one passes from $z$ to the complex variable
\beq
\rho=\frac{z}{(1+\sqrt{1-z})^2}\,.
\eeq
This variable was recently used in Ref.~\cite{Pappadopulo:2012jk} to study convergence of the conformal block decomposition. As we showed here, this is also an ideal variable for constructing rapidly convergent expansions of the conformal blocks themselves. The expansion coefficients $B_{n,j}$ satisfy a recursion relation which is a bit more involved than for the $A_{n,j}$: the coefficients at level $n$ are linear combinations of coefficients at all levels $n-2$, $n-4,\ldots$ up to zero, while for $A_{n,j}$ only the level $n-1$ contributes. But this complication pays off: the resulting coefficients do not exhibit any growth with $\Delta$ or $l$. This means that the coefficients computed and stored up to some large and fixed level $N$ can be used to evaluate conformal blocks of arbitrary dimension and spin with uniform accuracy.

Amazingly, even the first term in these infinite $\rho$-series expansions provides already a pretty good approximation (within a few \%) to the full conformal block. We use this fact in section \ref{sec:toy} to propose the ``toy bootstrap equation". Although this equation discards some information compared to using the full conformal blocks, and hence is less constraining, it has an advantage of involving only elementary function and being amenable to analytic analysis. Using this ``toy bootstrap", we get analytic understanding of why the method of \cite{Rattazzi:2008pe} was able to get an upper bound on the lowest dimension in the OPE. 

We believe that our $\rho$-series expansions will find many other future uses in the bootstrap program. Some of the possibilities are described in section \ref{sec:appl}. We conclude here by considering another possible application: bootstrap analysis of four point functions of non-scalar external primaries. Conformal blocks for such correlators have been studied recently in \cite{Costa:2011dw,SimmonsDuffin:2012uy}\footnote{The 2$d$ case was analyzed exhaustively in \cite{Osborn:2012vt}.}, but the results were not yet put to concrete use. Partly this is due to the fact that the obtained expressions are still rather complicated and not fully general. For example, Ref.~\cite{Costa:2011dw} finds only the blocks corresponding to the symmetric traceless exchanged primaries, while more general representations can be exchanged if the external fields have spin. We believe that the $\rho$-series approach could be useful in the problem of expressing these missing conformal blocks. The basic building blocks will no longer be simple Gegenbauers, but they will still be polynomials of the angular variable, fixed by the $SO(d)$ group theory. Once the expansion basis is known, the coefficients can presumably be found by using the Casimir equation judiciously.
It would be interesting to carry out this computation in detail.

\section*{Acknowledgements}
We are grateful to Balt van Rees for useful discussions about Zamolodchikov's variable, and to High Osborn and David Poland for valuable comments on the draft.
The work of S.R. is supported in part by the European Program ÒUnification in the LHC EraÓ, 
contract PITN-GA-2009-237920 (UNILHC).



\appendix

\section{Boundedness of the $\rho$-series coefficients }
\label{sec:largeL}

In this appendix we show that the coefficients $B_{n,j}$ on each level $n$ are uniformly bounded for all $\Delta$ and $l$, as stated in Eq.~(\ref{eq:bounded}). We have already shown in the main text that $B_{n,j}$ remain bounded as $\DD \to \infty$ for each fixed $l$. So here it suffices to consider the case of $l$ large with respect to $n$, say $l\ge n$. 

We will proceed by induction, and assume that the inequality has already been shown for all levels $n'<n$. Using Eq. (\ref{eq:Brecursion}), the bound at level $n$ will follow if we show that
\beq
\label{eq:factors}
{\Gamma^{\DD+n,j}_{\DD+n',j'}}/(C_{\DD+n,j} - C_{\DD,\l})
\eeq
is bounded by a constant which depends only on $n$ and $\nu$.

Our first observation is that the $\Gamma$'s satisfy the bound
\beq
\left| \Gamma^{\DD+n,j}_{\DD+n',j'}\right| \le const. \Delta+const 
\label{eq:Gboound}
 \eeq
 with constants which depends only on $n$ and $\nu$. To show this, notice that large contributions to $\Gamma$'s appear from only two sources. First, through the action of $r\del_r$, which gives a factor of $(\Delta+n')$. Second, through the action of $2\eta (1-\eta^2) \pd_\eta$, which gives factors $b_{j'}=O(j')$. On the other hand, all the factors produced via the expansion of denominators in \reef{eq:Dtilde} will depend only on $n$ and $\nu$. Notice in particular that $a_j=O(1)$.

Passing to the Casimir difference in (\ref{eq:factors}), we write it as
\beq 
C_{\DD+n,j} - C_{\DD,\l} = k^2 + 2k(l-n+\nu) +2n(\tau +n-1)\,,
\label{eq:CasDiff}
\eeq
where $j = l -n+ k$, $k = 0, 2, \ldots, 2n,$ and $\tau=\Delta-l-2\nu\ge0$ by the unitarity bounds. Since we are assuming $l\ge n$, this is a manifestly monotonically increasing function of $k$ and $n$. 

Consider first the case $k\ge 2$. In this case we have a lower bound:
\beq
\label{eq:bound_l}
C_{\DD+n,j} - C_{\DD,\l} \geq [\text{\reef{eq:CasDiff} for }k=n=2]=4(\Delta-\nu)\,\qquad (k\ge 2)\,.
\eeq
Combining this with \reef{eq:Gboound}, we see that \reef{eq:factors} is indeed bounded independently of $\Delta$ and $l$,
except in the region near the free scalar unitarity bound $\Delta=\nu$, excluded from consideration as discussed in the main text.

It remains to consider the case $k = 0$, when the Casimir difference
\beq 
C_{\DD+n,\l-n} - C_{\DD,\l} = 2n(\tau +n-1)\,
\label{eq:CasDiff2}
\eeq
can remain small even though both $\DD$ and $\l$ become large. However, precisely in this case the bound \reef{eq:Gboound} can also be improved. The relevant recursion relation coefficients are 
\beq
\Gamma^{\DD+n,\l-n}_{\DD+n',j'=\l-n'}\,.
\label{eq:lowestCoef}
\eeq
The coefficients with $j'\ne\l-n'$ will be zero, because lowering the spin via Eqs.~\reef{eq:idgeg} is accompanied by raising the dimension by at least the same amount. We will now show that all coefficients of the form (\ref{eq:lowestCoef}) are bounded 
by $const. \tau + const.$ Together with \reef{eq:CasDiff2}, this will prove the boundedness of $B_{n,\l-n}$ and will complete the proof.

For the case $n-n'=2$, this stronger bound can already be suspected in the expression (\ref{eq:G2}) for $\Gamma_{E,j}^{E+2,j-2}$, which contains two near-canceling terms. In detail, this coefficient can be expressed as
\beq 
\label{eq:Gamma_0}\Gamma_{\DD+n',\l-n'}^{\DD+n'+2,\l-n'-2} = 4 (\tau + 2n') a^{-}_{\l-n'}\,, 
\eeq
and satisfies the claimed bound, since $a^{-}_j \leq 1/2$ for all $j$.

For the general case, we notice that the action of $\widetilde{\mca D}$ on $\mca{P}_{\DD+n',\l-n'}$ can be written as follows:
\beq
\label{eq:expandD}
\widetilde{\mca{D}} \mca{P}_{\DD+n',\l-n'} = -\frac{4(\tau + 2n') a^{-}_{\l-n'}}{1-2 r^2(2 \eta^2-1)} \mca{P}_{\DD+n'+2,\l-n'-2} + \ldots
\eeq
Here we computed explicitly the action of  $(1-2\eta^2) r\del_r$ and $2\eta (1-\eta^2) \pd_\eta$. We omitted many terms $(\ldots)$ which cannot contribute to the relevant $\Gamma$ coefficients, because they raise the dimension without lowering the spin by the same amount. The $\Gamma$ coefficients \reef{eq:lowestCoef} with $n - n'= 4, 6, \ldots$ are obtained by expanding the denominator in (\ref{eq:expandD}). They are given by
\beq 
\Gamma_{\DD+n',\l-n'}^{\DD+n,\l-n} = 2 (\tau + 2n') \prod_{j=l-n',l-n'-2,\ldots,l-n+2}(2a^{-}_{j})\,,
\eeq
and clearly satisfy the claimed bound.

The reader will have noticed that the coefficients $B_{n,l-n}$, which required a separate analysis in the above proof, satisfy a recursion relation among themselves. This is because the relevant $\Gamma$'s in \reef{eq:lowestCoef} vanish for $j'\ne l-n'$. Due to this fact, these coefficients can in fact be computed explicitly:
\beq
\label{eq:Bexplicit}
B_{2m,l-2m} = \frac{(1/2)_{m}}{m!} \frac{(l+1-2m)_{2m}}{(\l+\nu+1-2m)_{2m}} \frac{(\tau/2)_m}{ (\tau/2+1/2)_m}\,.
\eeq
Their boundedness is also obvious from this formula.

\bibliographystyle{utphys}
\bibliography{Rad-biblio}

\end{document}